\documentstyle[aps,prl,multicol]{revtex}

\begin{document}

\input psfig.sty

\title
{\bf Measurement of the Proton's Neutral Weak Magnetic Form Factor}

\author{
B. Mueller$^{1}$, D. H. Beck$^{2}$, E. J. Beise$^{3}$, E. Candell
\footnote{Deceased}$^{4}$, L. Cardman\footnote{Present
address:  Thomas Jefferson 
National Accelerator Facility, Newport News, VA 23606}$^{2}$, R. Carr$^{1}$,  
R. C. DiBari$^{4}$, G. Dodson$^{5}$, K. Dow$^{5}$, F. Duncan$^{3}$, 
M. Farkhondeh$^{5}$, 
B. W. Filippone$^{1}$, T. Forest$^{2}$, H. Gao\footnote{Present address:
Physics Division, Argonne National Laboratory, Argonne, IL 60439}$^{2}$,
W. Korsch\footnote{Present address:  Department of Physics and
Astronomy, University of Kentucky, Lexington, KY  40506}$^{1}$, 
S. Kowalski$^{5}$, A. Lung$^{3}$, R. D. McKeown$^{1}$, R. Mohring$^{3}$,
J. Napolitano$^{4}$, D. Nilsson$^{2}$, M. Pitt\footnote{Present
address:  Department of 
Physics, Virginia Polytechnic Institute and State University, Blacksburg, 
VA 24061-0435}$^{1}$,
N. Simicevic\footnote{Present address:  Department of Physics, 
Louisiana Tech University, 
Ruston, LA 71270}$^{2}$, B. Terburg$^{2}$,
and S. P. Wells$^{\dag\dag 5}$
}

\address{
$^{1}$ Kellogg Radiation Laboratory, California Institute of Technology
Pasadena, CA 91125, USA \\
$^{2}$ University of Illinois at Urbana-Champaign, Urbana, Illinois
61801 \\
$^{3}$ University of Maryland, College Park, Maryland 20742 \\
$^{4}$ Rensselaer Polytechnic Institute, Troy, New York 12180 \\
$^{5}$ Bates Linear Accelerator Center, Laboratory for Nuclear Science
and Department of Physics,\\
Massachusetts Institute of Technology, Cambridge, Massachusetts 02139
}
\maketitle
\vskip 12pt
\centerline{(SAMPLE Collaboration)}
\vskip 12pt
\date{\today}

\begin{abstract}
We report the first measurement of the parity-violating asymmetry in elastic
electron scattering from the proton. The asymmetry depends on
the neutral weak magnetic form factor of the proton which contains
new information on the contribution of strange quark-antiquark pairs 
to the magnetic moment of the proton. We obtain the value 
$G_M^Z= 0.34 \pm 0.09 \pm 0.04 \pm 0.05$ n.m. at $Q^2=0.1$ (GeV/c)${}^2$. 
\end{abstract}

\begin{multicols}{2}[]
\narrowtext
\vfill
\eject
The measurement of strange quark-antiquark ($\bar s s$) effects in 
the nucleon offers
a unique window to study the effects of the $\bar q q$ ``sea''  
at low momentum transfers. This information is an important clue to
the dynamical effects of QCD that are responsible for form factors
in the non-perturbative regime, and may lead to new insight into the
origins of these effects.

It has been shown${}^1$
that the neutral weak
current can be used to determine the  
$\bar s s$ contributions to nucleon form factors.
The magnetic moment is one important nucleon property that
can be studied in this fashion. The
neutral weak magnetic form factor of the proton can be measured in 
parity-violating electron scattering,${}^2$  thus providing information on
the $\bar s s$ content 
of the nucleon's magnetic moment. 
In this Letter, we report the first such measurement and obtain the
first direct experimental data relevant to determination of the strange 
magnetic moment of the proton.

To lowest order (tree-level),
the neutral weak magnetic form factor of the proton $G_M^Z$ can be related to 
nucleon 
electromagnetic form factors and a contribution from strange quarks:
\begin{eqnarray}  
G_M^Z = {1 \over 4} (G_M^p-G_M^n) - \sin^2\theta_W \> G_M^p - {1 \over 4}
G_M^s
\end{eqnarray}  
where $G_M^p$ and $G_M^n$ are the (electromagnetic) nucleon magnetic form
factors, and
$\theta_W$ is the weak mixing angle. 
(Note that the weak mixing angle has recently
been determined${}^3$ with high precision: 
$\sin^2 \theta_W (M_Z) = 0.2315 \>.$)
Electroweak radiative corrections must be applied 
to the coefficients in Eq. (1), which have been computed in Ref. 4. 
Then measurement 
of the neutral weak form factor $G_M^Z$ will allow (after combination with the
well known electromagnetic form factors) determination of the strange
magnetic form factor $G_M^s$. There have been a variety of theoretical 
predictions for $G_M^s$ in the limit of zero momentum transfer
($\equiv \mu_s$) over the last few
years, and summaries of them are presented in references $5$ and $7$. The 
typical magnitude is $-0.3$ n.m., although the predictions range from
-0.73 to +0.42 n.m. .

As mentioned above, the quantity $G_M^Z$ for the proton can be measured
via elastic parity-violating electron scattering at backward angles.${}^2$
The difference in cross sections for right and left handed incident
electrons arises from interference of the electromagnetic and
neutral weak amplitudes, and so contains products of electromagnetic
and neutral weak form factors. The expression for elastic
scattering from the proton is given by
\end{multicols}

\widetext
\begin{eqnarray}
A& = &{\sigma_R - \sigma_L}\over {\sigma_R + \sigma_L}\hfil \nonumber \\
 & = &\left[- G_F Q^2 \over  \pi \alpha \sqrt{2}\right] 
{{ 
\varepsilon G^{\gamma}_{{E}} G^{Z}_
{{E}} + \tau G^{\gamma}_{{M}}
G^{Z}_{{M}} - {1 \over 2}(1-4 \sin^2 \theta_W ) 
\varepsilon^{\prime} G^{\gamma}_{{M}} G^{Z}_{A}}   \over
{\varepsilon (G_
{{E}})^2 + \tau (G_{{M}})^2}}
\end{eqnarray}
\begin{multicols}{2}[]
\narrowtext

\noindent where $ \varepsilon $, $\varepsilon^{\prime}$, and $\tau$
are kinematic quantities, and $Q^2>0$ is the four-momentum transfer.${}^5$
In addition to
the electric and magnetic neutral weak form factors, $G_E^Z$ and $G_M^Z$,
the numerator of this expression also contains the 
neutral weak axial form factor:
\begin{eqnarray}
G_A^Z = -{1 \over 2} (1 + R_A) G_A + {1 \over 4} G_A^s
\end{eqnarray}

\noindent where we take $G_A = 1.2601$ from neutron beta decay${}^3$,
$G_A^s= \Delta s \sim -0.1$
is from polarized deep inelastic scattering${}^6$, and
$R_A \sim -0.34$ is the axial radiative correction.${}^{4,5}$ 
This axial radiative correction is quite uncertain and we assign a 100\%
uncertainty to this correction. The uncertainty due to 
the $G_A^s$ term is about an order of magnitude smaller than the radiative
correction uncertainty.

The SAMPLE experiment measures the parity-violating
asymmetry in the elastic scattering of 200 MeV incident electrons at 
backward angles with an average $Q^2 \simeq
0.1$(GeV/c)${}^2$
For $G_M^s=0$, the value of $G_M^Z$ 
(at $Q^2 = 0.1$ GeV${}^2$) is 0.40 and the expected asymmetry in 
the SAMPLE experiment is $-7.2 \times 10^{-6}$ or 
-7.2 ppm; for $G_M^s =-0.3$ n.m.,  $G_M^Z = 0.48$
and we would then expect $A = -8.3$ ppm.
At these
kinematics the axial term contributes about 30\% 
of the asymmetry, so the uncertainty due to the axial
term is $\pm 0.7$ ppm.

The experiment is performed at the MIT/Bates Linear Accelerator Center
using a 200 MeV polarized electron beam 
incident on a liquid hydrogen target. The scattered electrons
are detected in a large solid angle ($\sim 2$ sr) Cerenkov detector at backward
angles $130^\circ < \theta < 170^\circ$ corresponding to the range of
momentum transfers $0.10 < Q^2 <
0.11$ (GeV/$c)^2$.

The liquid hydrogen target is 40 cm long, and is part of a high flow-rate
recirculating liquid hydrogen system with a heat exchanger to remove the
$\sim 500$ watts of heat deposited by the electron beam.
Studies of the 
performance${}^8$ of the target indicate that with 40 $\mu$A of beam
the target can be maintained as a sub-cooled liquid and that density
fluctuations are less than 0.1\%.
The fluctuations in the detected signal are then dominated by
counting statistics of the scattered particles.

The detector consists of 10 large mirrors, each with ellipsoidal curvature
to focus the Cerenkov light onto 
one of ten photomultiplier tubes. Each photomultiplier is 8 inches in diameter
and is shielded from the target and room background by a cast lead shield.
In addition, 2mm of Pb
on the scattering chamber wall is essential to reduce the x-ray
background in the detector to a reasonable level.

A remotely controlled light
shutter can cover each photomultiplier tube for background
measurements. We typically take one third
of our data with shutters closed to monitor this background.
We have determined that this background is soft electromagnetic 
radiation (a thin Pb shield
eliminates it) and arises from showering in the target plus additional
scattering from downstream of the apparatus. Electromagnetic background 
generated by soft bremsstrahlung will have negligible helicity dependence.
(The parity-violating asymmetry is proportional to the squared momentum 
transfer to the proton which is extremely small.) Our measurements do
verify an absence of helicity dependence to this signal (see Table 1). 
In addition, we study the detector signals with covers on the mirrors
to analyze the signals we observe for non-Cerenkov sources of light. 
The observed elastic scattering signal and background levels are
in good agreement with expectations. 

The incident electron beam is pulsed at 600 Hz; 
each detector signal is integrated over the
$\sim 15 \mu$sec of every beam pulse and digitized. The beam intensity
is similarly integrated and digitized. The ratio of integrated detector signal
to the integrated
beam signal is the normalized yield which is proportional to the cross section
(plus background).
We then measure the helicity dependent asymmetry in the normalized yield which
yields the parity-violating asymmetry $A$. All
10 detectors are combined in software during the data analysis.

The polarized electron source is a GaAs photoemission source.
The laser beam that is incident on the 
GaAs crystal is circularly polarized by a $\lambda /4$ Pockels cell.
The electron beam helicity is rapidly reversed by changing the voltage 
on the Pockels
cell to reverse the circular polarization of the light. The
helicity is randomly chosen for each of 10 consecutive beam pulses
and then the complement helicities are
used for the next 10 pulses. 
The asymmetry in the normalized yields 
is computed for
``pulse pairs'' separated by 1/60 of a second to minimize systematic
errors. Each pulse pair asymmetry corresponds to a measurement 
of the parity-violating cross section asymmetry.
These are then
combined in 30 minute runs corresponding to typically 0.5 million measurements.
An active feedback system${}^9$ keeps the
helicity correlations in the beam intensity averaged to zero to better than
1 ppm.

The beam helicity can be manually reversed
by rotating a $\lambda$/2 plate which reverses the helicity of the light
(and the beam) relative to all electronic signals. We denote this setting
as ``reverse'' as opposed to the ``normal'' setting;
a real parity violation
signal will appear to change sign under this ``slow reversal''. Electronic
crosstalk and other effects will not change under ``slow reversal'', 
so this is an important test that our signal is not some spurious 
systematic effect.

The electron polarization is measured using a Moller apparatus on the
beamline and is typically 35\%. A Wien spin rotator located at the
exit of the polarized source is used to rotate the spin for investigation
of transverse spin asymmetries and minimize their effects during 
parity violation measurements. Based on these studies, the effect of 
transverse components of
electron polarization to the observed parity violation signal is determined 
to be negligible.

The beam position and angle at the target in both transverse
dimensions ($x$ and $y$), the beam energy, and the ``halo'' of the beam
are continuously monitored for every beam pulse. If a helicity correlation
is present, then we can correct the detector normalized yield
asymmetries to remove its effect. 
In practice, these corrections are generally less than
1ppm per run and approximately average to zero over many runs.
Other properties, such as the width of 
the beam,
are studied during special runs to verify the absence of helicity correlations.

An important consideration in analyzing the data is the fraction of
detector signal that is actually from elastic scattering. The
breakdown of our signal is summarized in Table 1 along with
information on the asymmetries associated with all background
processes.
In order to study these backgrounds, we developed a technique of accelerating
``tracer bullets'' of $\sim 0.1$mA peak current interspersed with
very weak beam pulses with typical peak currents in the nA range.
One can then use more conventional pulse counting techniques to study
the Cerenkov detector response. This method clearly shows the relative
fraction of scintillation light (events with only one photoelectron)
relative to Cerenkov light (events with several photoelectrons). 
The Cerenkov
light is almost totally due to elastic scattering of electrons from protons,
with a small contribution due to pion decays. The pion background is rather 
small because the low incident beam energy (200 MeV) is close to 
the production threshold. The pion
yield has been calculated using known photoproduction cross sections, and 
the low $\pi^+$ yield is verified by measurement of the rate of delayed 
signals from muon decays just after the electron beam pulse ends.   
In addition, we use a NaI detector
behind a mirror in coincidence to check the spectrum of scattered electrons
and compare with computer simulations.

We have acquired a significant data sample from two runs; one in fall 1995 
and one in spring 1996.
Combining the data in each of the 
two data sets
and computing the physics asymmetry, we obtain the asymmetries
shown in Figure 1. 
The corrections to the raw measured asymmetry are
the background effects given in Table 1 ($1.82 \pm 0.13$), the
correction for (internal and external) bremsstrahlung radiation 
($1.16 \pm 0.01$), 
and the beam polarization ($1/(0.348 \pm 0.015)$). These are combined to
give an overall multiplicative correction of $6.07 \pm 0.51$.
We have assumed that the shutter closed background and the x-ray background
light have zero asymmetry, as expected for soft electromagnetic processes.
Combining the asymmetries from the two data sets yields the
value for the parity-violating asymmetry
$$
A = -6.34 \pm 1.45 \pm 0.53 \> {\rm ppm}
$$ 
where the first uncertainty is statistical and the second is the estimated
systematic error. The neutral weak magnetic form factor derived
from this asymmetry is
$$
G_M^Z(Q^2=0.1 {\rm GeV}^2) = 0.34 \pm 0.09 \pm 0.04 \pm 0.05 \> {\rm n.m.}
$$
where the last uncertainty is due to the axial radiative
correction and other minor form factor uncertainties associated with 
extracting $G_M^Z$ from the measured asymmetry.
In the absence of a strange quark contribution we expect
$G_M^Z = 0.40$ n.m.;
our measurement of $G_M^Z$ thus corresponds to a value of 
$G_M^s (Q^2 = 0.1 {\rm GeV}^2) = +0.23 \pm 0.37 \pm 0.15 \pm 0.19 \> 
{\rm n. m.}$

If we do not assume the shutter closed background to be
identically zero but rather use the measured shutter closed
asymmetry (see Table 1), then the
result for $A$ becomes $-7.38 \pm 2.12 \pm 0.66$ ppm where the statistical
error is increased by the shutter closed measurement. This value is
in good agreement with that obtained above.

In the future,
for the full hydrogen data set we project that the statistical error on
$G_M^Z$ will be reduced to about 0.05 n.m.
Additional running with deuterium will further reduce the 
uncertainty in $G_M^Z$ by greatly reducing the error associated with 
the axial radiative correction.${}^{11}$ The deuterium asymmetry
contains the same fraction of axial correction but is almost
completely insensitive to $G_M^s$.
As a result the ratio $A_p/A_D$ has the same 
sensitivity to $G_M^s$ but almost no sensitivity to $R_A$.

Additional experiments to explore other features of neutral weak currents
and strange form factors of the nucleon are planned at Mainz (MAMI-B)${}^{12}$
and TJNAF (formerly known as CEBAF)${}^{13}$. 
These experiments will provide a new
and unique window on the quark structure of the nucleon, and hopefully
will provide important information towards a more complete understanding
of nucleon structure in the context of QCD.

The persistent efforts of the staff of the MIT/Bates facility to provide
high quality beam and improve the experiment are gratefully acknowledged.
Financial support for the construction of the Moller apparatus from CEBAF
is acknowledged.
This work was supported by
NSF grants PHY-9420470 (Caltech), PHY-9420787 (Illinois), 
PHY-9457906/PHY-9229690 (Maryland), PHY-9507412 (RPI), and DOE
cooperative agreement DE-FC02-94ER40818 (MIT-Bates).

\begin{figure}
\centerline{\psfig{figure=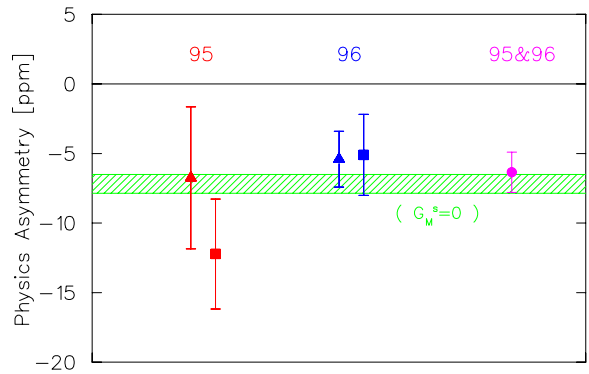,angle=90,width=3.0in}}
\vskip 0.2 in
\caption{Results for the parity-violating asymmetry
measured in the 1995 and 1996 running periods. For each running period
we display the result for the ``normal'' (squares) and ``reverse'' (triangles)
settings of slow helicity reversal, which are in good agreement. The
combined result for both helicity states and both running periods is 
also shown (circle).
The error bars include
statistical errors only.
The hatched region indicates the asymmetry band (due to the uncertain axial
radiative correction) for $\mu_s = G_M^s =0$.}
\end{figure}

\begin{table}
\begin{center}
\caption{Components of Detector Signal}
\begin{tabular}{l|c|c|} \hline
Source & Fraction & Asymmetry \\ \hline
Elastic electrons & 54.8 $\pm$ 4.0\% & $A$ \\
Shutter closed & 22.6 $\pm$ 0.5\% & 1.95 $\pm$ 3.13 ppm (measured) \\
$\pi$ decays & 2.3 $\pm$ 1.4\% & $<$ 1 ppm (Ref. 10) \\
X-rays (EM) & $20.3 \pm 3.7\%$ & 0 \\ \hline 
\end{tabular}
\end{center}
\end{table}
\end{multicols}
\end{document}